\documentclass[aps,prd,showpacs,twocolumn,
10pt,superscriptaddress,floatfix,nofootinbib,longbibliography]{revtex4-1}
\usepackage{amssymb,amsfonts,amsmath,bm,bbm}   
\usepackage{graphicx}
\usepackage{dcolumn}
\usepackage{bm}
\usepackage[utf8]{inputenc}
\usepackage[T1]{fontenc}
\usepackage{minted}  
\usepackage[
colorlinks=true,
linkcolor=blue,
breaklinks=true,
urlcolor=blue,
citecolor=blue]{hyperref}
\usepackage{bbold}
\usepackage{epstopdf}

\newcommand{\HNUST}{\affiliation{
Hunan Provincial Key Laboratory of Intelligent Sensors and Advanced Sensor Materials, \\ School of Physics and Electronics, Hunan University of Science and Technology, Xiangtan 411201, China}} 

\newcommand{\HNUSTchem}{\affiliation{
School of Chemistry and Chemical Engineering, Hunan University of Science and Technology, Xiangtan 411201, China}}

\newcommand{\LZU}{\affiliation{
School of Nuclear Science and Technology, Lanzhou University, Lanzhou 730000, China}}

\newcommand{\UOC}{\affiliation{
Department of Physics and Astronomy, University of Catania, Via Santa Sofia 64, 1-95125 Catania, Italy}} 

\newcommand{\INFN}{\affiliation{
INFN-Sezione di Catania, Via Santa Sofia 64, I-95123 Catania, Italy}}

\begin{document}

\title{
From lattice QCD thermodynamics to an effective quasiparticle description of gluonic matter at finite temperature
}

\author{Cheng-Gang Guo}
\HNUST

\author{La-Yuan Peng}
\HNUST

\author{Zhen-Yan Lu}
\email{luzhenyan@hnust.edu.cn}
\HNUST

\author{Marco Ruggieri}
\email{marco.ruggieri@dfa.unict.it}\UOC\INFN

\author{Shu-Peng Wang}\LZU\HNUST

\author{Zhi-Jun Ma}
\email{mazhijun@hnust.edu.cn}
\HNUST\HNUSTchem

\begin{abstract}

We develop a lattice-constrained quasiparticle description of pure SU(3)
Yang--Mills matter at finite temperature. The temperature-dependent
effective gluon quasiparticle mass is determined from the lattice
entropy density over $0.70\leq T/T_c\leq5$. It is interpreted as an
effective thermal parameter encoding medium modifications of gluonic
excitations, rather than as a physical gluon mass. With this mass fixed,
we reconstruct the background contribution required for thermodynamic
consistency, using a single pressure-normalization condition. The
resulting pressure, energy density, entropy density, interaction
measure, and squared speed of sound are then compared with available
lattice results. Near $T_c$, the rapid variation of the effective mass is accompanied by
a pronounced maximum in the interaction measure at $T/T_c\simeq1.09$,
a strong suppression of the squared speed of sound, and a maximum in
the background contribution. These correlated features characterize
the substantial departure from conformal behavior in the transition
region. At higher temperatures, the interaction measure decreases and
the squared speed of sound approaches the conformal limit. We also
examine the energy-per-excitation ratio $E/n_g$, which has a minimum of
approximately $964~\mathrm{MeV}$ at $T/T_c\simeq1.21$. This ratio
provides a measure of the excitation scale within the quasiparticle
description. Overall, our results show how lattice thermodynamics can
be mapped onto an effective gluonic excitation spectrum while
maintaining thermodynamic consistency.

\end{abstract}

\maketitle


\section{Introduction}

The thermodynamics of strongly interacting matter at finite temperature
provides an important connection between the microscopic dynamics of
quantum chromodynamics (QCD) and the collective properties of matter
under extreme conditions
\cite{Gross:2022hyw,Brambilla:2014jmp,MUSES:2023hyz,Gelis:2021zmx,
Chen:2024aom,Altmann:2024icx}. 
Asymptotic freedom implies that QCD becomes weakly coupled at sufficiently
high momentum scales, whereas confinement governs its infrared dynamics
\cite{Gross:1973id,Gross:1973ju,Politzer:1973fx,DeTar:2009ef,Li:2025uaw}.
At finite temperature, the interplay between these regimes gives rise to
deconfined color degrees of freedom at high temperature
\cite{Polyakov:1978vu,Susskind:1979up,Svetitsky:1982gs,
Yaffe:1982qf,Aarts:2023vsf,Harris:2023tti,Bazavov:2009zn}.
Pure $SU(3)$ Yang--Mills theory provides a particularly clean setting for
studying this thermal dynamics. In the absence of dynamical quarks, the $Z_3$ center symmetry of the 
$\mathrm{SU}(3)$ Yang--Mills theory is exact, and the Polyakov loop serves as 
an order parameter for the deconfinement transition. Lattice simulations have 
established the first-order nature of this transition at the critical 
temperature $T_c$ 
\cite{Engels:1990vr,Boyd:1995zg,Boyd:1996bx,Shirogane:2016zbf,
Ejiri:2019csa,Lucini:2023irm}. Below $T_c$, the physical asymptotic spectrum 
is composed of color-singlet states, with glueballs providing the 
lowest-lying excitations 
\cite{Bali:1993fb,Morningstar:1999rf,Chen:2005mg,Athenodorou:2020ani}. 
Nevertheless, quasiparticle-based descriptions have been widely employed as 
effective approaches to characterize the thermodynamic properties of strongly 
interacting matter, including the confined region. In such frameworks, 
gluonic quasiparticles need not be interpreted as freely propagating colored 
states, but rather as effective thermodynamic variables that encode medium effects,
including those associated with confinement. These effects can be 
incorporated through infrared-modified dispersion relations, Polyakov-loop 
backgrounds, or strong quasiparticle interactions 
\cite{Islam:2021qwh,Lo:2013hla,Filinov:2014dda,Heffner:2012sx,
Kang:2022jbg,Kang:2024xqk}. Alternative descriptions of the low-temperature 
regime have also been developed using color-singlet glueball quasiparticles 
or Polyakov-loop-modified effective excitations 
\cite{Hell:2008cc,Buisseret:2010mop}. In the present work, the gluonic 
quasiparticles below $T_c$ are therefore regarded as effective variables for 
reconstructing the thermodynamic equation of state rather than as physical 
colored asymptotic excitations. 

Lattice gauge theory provides the most reliable first-principles
determination of the equation of state of pure-glue matter. Early
calculations established the characteristic behavior of the pressure and
energy density across the transition
\cite{Engels:1990vr,Boyd:1995zg,Boyd:1996bx}, while subsequent studies
substantially improved the control of lattice-spacing, finite-volume, and
continuum-extrapolation effects
\cite{Borsanyi:2012ve,Francis:2015lha,Giusti:2016iqr,
Shirogane:2016zbf,Kiyohara:2021smr,Borsanyi:2022xml}.
These calculations reveal pronounced thermodynamic variations near $T_c$
and sizable deviations from ideal-gas behavior over a broad temperature
range. In particular, the interaction measure is strongly enhanced around
the deconfinement region, while the squared speed of sound is substantially
suppressed relative to its conformal value
\cite{Gavai:2005dy,Karsch:2008fe,Datta:2009jn}.
At sufficiently high temperatures, weak-coupling and resummed approaches
become increasingly applicable
\cite{Braaten:1989mz,Andersen:1999fw,Andersen:1999sf,
Andersen:2002ey,Kajantie:2002wa,Kajantie:2003ax,Laine:2006cp,
Andersen:2010ct,Strickland:2010tm,dEnterria:2022hzv}, whereas the region
near $T_c$ remains strongly affected by nonperturbative dynamics. This provides a natural motivation for effective descriptions capable of incorporating the substantial medium modifications seen in lattice thermodynamics.

Quasiparticle models implement this idea by representing the interacting
medium in terms of thermal excitations with medium-dependent dispersion
relations, where interaction effects are effectively encoded through
quantities such as temperature- or chemical-potential-dependent quasiparticle
masses \cite{Peshier:1995ty,Levai:1997yx,Plumari:2011mk,Li:2022ozl,Castorina:2011ja,Castorina:2011ra}. Early quasiparticle descriptions of hot gluonic matter were introduced by
Goloviznin and Satz and subsequently developed into thermodynamically
self-consistent formulations \cite{Goloviznin:1992ws,Peshier:1994zf,Gorenstein:1995vm}.
The framework has since been extensively applied to pure-glue and full-QCD
thermodynamics \cite{Peshier:1999ww,Schneider:2001nf,Peshier:2002ww,
Bluhm:2004xn,Alba:2014lda,Soloveva:2019xph,Mykhaylova:2020pfk},
while more recent studies have addressed hot and dense matter,
momentum-dependent quasiparticle properties, transport phenomena, and
lattice-constrained extractions of effective partonic properties
\cite{Ma:2023stj,Pal:2024nza,Sambataro:2024mkr,Soloveva:2023tvj,Jamal:2024qzy}. 
It is important to emphasize that the quasiparticle mass introduced in such
approaches is not a bare gauge-boson mass nor a physical pole mass of a
propagating gluon. Instead, it serves as an effective thermodynamic parameter
that parametrizes the medium-induced modification of thermal gluonic
excitations. In particular, when the quasiparticle description is extended
into the confined region below $T_c$, the large effective mass reflects the
suppression of colored thermal excitations required to reproduce the lattice
thermodynamics, rather than the existence of freely propagating massive gluon
states \cite{Peshier:1995ty,Levai:1997yx}.

When the quasiparticle mass depends explicitly on temperature, however,
thermodynamic differentiation generates additional contributions and
requires a consistent treatment of the medium dependence
\cite{Gorenstein:1995vm,Peshier:1999ww}.
Within the Gorenstein--Yang-type formulation, thermodynamic consistency is
restored by introducing a temperature-dependent background contribution
whose variation compensates for the implicit temperature dependence of the
quasiparticle spectrum
\cite{Gorenstein:1995vm,Peshier:1999ww,Schneider:2001nf,
Peshier:2002ww,Bluhm:2004xn}.
Because this background contribution enters the pressure and energy density
with opposite signs, it cancels from the enthalpy density
\cite{Gorenstein:1995vm,Peshier:1999ww}.
At vanishing chemical potential, the entropy density therefore provides a
particularly direct constraint on the quasiparticle sector, making lattice
entropy data especially useful for determining effective thermal
quasiparticle properties
\cite{Bluhm:2004xn,Plumari:2011mk,Li:2022ozl,Soloveva:2023tvj}.

In this work, we construct a lattice-constrained thermodynamically consistent
quasiparticle description of pure $SU(3)$ Yang--Mills matter over the range
$0.70\leq T/T_c\leq5$. The central idea is to use the lattice entropy density
as the sole thermodynamic input for determining the temperature-dependent
effective gluon mass $m_g(T)$. Once this entropy constraint fixes the
quasiparticle spectrum, the background contribution required by thermodynamic
consistency is uniquely reconstructed from the corresponding consistency
relation with only a single pressure normalization condition. Consequently,
the pressure, energy density, interaction measure, and thermodynamic response
functions emerge from the same lattice-constrained quasiparticle input without
introducing additional temperature-dependent fitting functions. 
This construction provides a minimal connection between lattice thermodynamics
and an effective description of thermal gluonic excitations. Unlike parametrizations based on perturbative assumptions or phenomenological inputs, the present approach directly extracts the effective excitation scale from lattice entropy density, providing a transparent connection between first-principles thermodynamics and quasiparticle properties. In particular, it
allows us to investigate how an effective quasiparticle spectrum constrained by lattice entropy density serves as an effective representation of medium-modified gluonic excitations while maintaining thermodynamic consistency. We further investigate the background contribution and the
isentropic bulk modulus to characterize the thermodynamic consistency and
mechanical response of pure gluon matter.

The paper is organized as follows. In Sec.~\ref{sec:II}, we formulate the
thermodynamically consistent quasiparticle model for pure $SU(3)$
Yang--Mills matter. In Sec.~\ref{sec:thermo}, the temperature-dependent
effective gluon mass is constrained by the lattice entropy density, and
the resulting equation of state and thermodynamic response are analyzed
and compared with the available lattice results. Section~\ref{sec:conclusion} summarizes
our main findings.

\section{Thermodynamically consistent quasiparticle model}
\label{sec:II}

The thermodynamics of the pure SU(3) Yang-Mills matter is investigated within a
thermodynamically consistent quasiparticle framework. In this approach, the
interacting gluonic medium is described in terms of thermal quasiparticles
characterized by a temperature-dependent effective mass $m_g(T)$
\cite{Goloviznin:1992ws,Peshier:1995ty,Schneider:2001nf}.
The corresponding modification of the quasiparticle dispersion relation
effectively incorporates medium-induced interaction effects into the
single-particle excitation spectrum. The effective gluon mass introduced here
should therefore not be interpreted as a bare gluon mass; rather, it
parametrizes the modification of gluonic excitations induced by the thermal
medium. Quasiparticle descriptions of this type have been widely employed to
reproduce lattice thermodynamics in the strongly interacting regime, where
straightforward perturbative treatments become inadequate
\cite{Levai:1997yx,Schneider:2001nf,Alba:2014lda}.

As discussed in the Introduction, a crucial issue in quasiparticle models is the preservation of thermodynamic consistency. When the quasiparticle mass depends explicitly on temperature, a direct substitution of $m_g(T)$ into the ideal-gas expressions leads to additional terms in thermodynamic derivatives. As a result, the pressure, entropy density, and energy density can no longer be obtained independently from the ideal-gas relations. As first emphasized by Gorenstein and Yang, a temperature-dependent background term is required to maintain the standard thermodynamic identities and ensure thermodynamic consistency \cite{Gorenstein:1995vm}. Similar self-consistent formulations have
subsequently been developed and applied to QCD thermodynamics
\cite{Bannur:2006hp,Ruggieri:2012ny,
Plumari:2011mk,Sasaki:2012bi,
Islam:2021dyk,Sasaki:2013xfa,Kang:2022jbg}.

Within this framework, the pressure of the pure gluon plasma is written as
\begin{equation}
P(T)=P_{\rm id}(T,m_g(T))-B(T),
\label{eq:P_qpm}
\end{equation}
where $P_{\rm id}$ denotes the ideal Bose-gas contribution evaluated with the
temperature-dependent effective gluon mass, while $B(T)$ is the background
term required by thermodynamic consistency
\cite{Gorenstein:1995vm,Peshier:1995ty}. The corresponding energy density is
\begin{equation}
E(T)=E_{\rm id}(T,m_g(T))+B(T).
\label{eq:E_qpm}
\end{equation}

For pure $SU(3)$ Yang--Mills theory, retaining the two transverse gluonic
polarization states gives the degeneracy factor $d_g=2(N_c^2-1)=16$. The quasiparticle dispersion relation is taken to be
\begin{equation}
\omega(k,T)=\sqrt{k^2+m_g^2(T)}.
\label{eq:dispersion}
\end{equation}
With this dispersion relation, the ideal-gas pressure and energy
density are given by
\begin{equation}
\begin{aligned}
P_{\rm id}
=
-\frac{d_gT}{2\pi^2}
\int_0^\infty
k^2 
\ln
\left(
1-e^{-\omega(k,T)/T} 
\right) dk ,
\label{eq:PidT4}
\end{aligned}
\end{equation}
and
\begin{equation}
\begin{aligned}
E_{\rm id}
&=
\frac{d_g}{2\pi^2}
\int_0^\infty
k^2 
\frac{
\omega(k,T)
}
{e^{\omega(k,T)/T}-1} 
dk .
\label{eq:EidT4}
\end{aligned}
\end{equation}

The background function $B(T)$ is determined by imposing the thermodynamic
consistency relation
\begin{equation}
\left(\frac{dP(T)}{dm_g}\right)_{T}=0,
\label{eq:s_derivative}
\end{equation}
together with the requirement that the entropy density retain its
quasiparticle form $S=dP/dT$. This yields
\begin{equation}
\frac{dB(T)}{dT}
=
\frac{\partial P_{\rm id}(T,m_g(T))}
{\partial m_g}
\frac{dm_g(T)}{dT}.
\label{eq:dB}
\end{equation}
Equivalently, using the Bose--Einstein distribution, Eq.~\eqref{eq:dB} can be written explicitly as
\begin{align}
\frac{dB(T)}{dT}
=&
-\frac{d_g}{2\pi^2}
\frac{dm_g(T)}{dT}m_g(T)
\nonumber\\
&\times
\int_0^\infty 
\frac{k^2}{\omega(k,T)}
\frac{1}{
e^{\omega(k,T)/T}-1
} dk.
\label{eq:dB_integral}
\end{align}
Equation~\eqref{eq:dB_integral} establishes a self-consistent relation between
the temperature-dependent quasiparticle mass and the background function. If
the functional form of the background contribution is specified, this
relation may be regarded as a differential equation for $m_g(T)$. Conversely,
once $m_g(T)$ is known, the background function can be obtained by direct
integration,
\begin{align}
B(T)
=&~
B(T_0)
-
\frac{d_g}{2\pi^2}
\int_{T_0}^{T} dT'
\frac{dm_g(T')}{dT'}m_g(T')
\nonumber\\
&\times
\int_0^\infty 
\frac{k^2}{\omega(k,T')}
\frac{1}{
e^{\omega(k,T')/T'}-1
} dk .
\label{eq:Bstar}
\end{align}
Here $T_0$ is a reference temperature and $B(T_0)$ is the corresponding
integration constant, to be fixed by an appropriate thermodynamic
normalization condition. Thus, once the temperature dependence of the
effective gluon mass is specified, the background contribution is completely
determined up to this single constant, and no additional
temperature-dependent function needs to be introduced. We emphases that a similar structure arises in thermodynamically consistent quasiparticle
descriptions of cold dense matter. At vanishing temperature, the medium
dependence is governed by the chemical potentials, so that the effective
quark masses and the corresponding background contribution generally take
the forms $m_i=m_i(\mu_i,g(\{\mu_j\}))$ and
$B=B(\{\mu_i\})$, respectively
\cite{Lu:2016fki,Qian:2023hrt}.
The running coupling may therefore introduce an additional implicit
chemical-potential dependence into the quasiparticle spectrum.
Thermodynamic consistency requires this medium dependence to be compensated
by the corresponding background contribution, in analogy with the
finite-temperature relation above. Related consistency considerations also
appear in modified perturbative-QCD descriptions of cold quark matter,
where an additional medium-dependent contribution to the thermodynamic
potential can be introduced to restore the thermodynamic identities~\cite{Xu:2014zea,Wang:2025lwv}.

An immediate consequence of Eqs.~\eqref{eq:P_qpm} and \eqref{eq:E_qpm} is
that the background contribution cancels from the enthalpy density, $E(T)+P(T)=E_{\rm id}(T,m_g)+P_{\rm id}(T,m_g)$. 
Accordingly, the entropy density becomes
\begin{equation}
S(T)
=\frac{E(T)+P(T)}{T}
=
S_{\rm id}(T,m_g(T)).
\label{eq:s_qpm}
\end{equation}
The cancellation of $B(T)$ in the entropy density is particularly useful in
the present framework, because it allows the effective quasiparticle mass to
be constrained directly by lattice thermodynamic information without an
explicit dependence on the background term.

In the present work, discrete values of the effective gluon mass are inferred
from the lattice equation of state~\cite{Borsanyi:2012ve}, with the entropy density serving as the
primary thermodynamic constraint. The resulting temperature dependence is
then represented by a smooth parametrization in terms of the dimensionless
variable $x=T/T_c$, as specified in the next section. The parametrization is
constructed to reproduce the rapid variation of the effective mass in the
vicinity of the deconfinement transition while providing a smooth continuation
into the high-temperature regime.

Once the lattice-constrained effective mass $m_g(T)$ has been determined,
the background function $B(T)$ and the complete thermodynamic equation of
state can be reconstructed self-consistently. In the following sections, we
investigate the temperature dependence of the effective gluon mass, pressure,
energy density, entropy density, background contribution, trace anomaly,
specific heat, squared speed of sound, and isentropic bulk modulus. We also
examine the energy and free energy per thermal quasigluon and compare the
corresponding thermodynamic observables with lattice results where available.


\section{Thermodynamic properties of the $SU(3)$ gluon plasma}
\label{sec:thermo}

Having established the thermodynamically consistent quasiparticle
framework, we now determine the remaining model input from lattice
thermodynamics and investigate the resulting properties of pure
$SU(3)$ Yang--Mills matter. The temperature-dependent effective gluon
mass $m_g(T)$ is first extracted from the lattice entropy density and
represented by a smooth analytic parametrization. The integration constant
associated with the background contribution is then fixed by a pressure
normalization condition at $T_c$. The complete equation of state follows
from the thermodynamic consistency relations without introducing additional
temperature-dependent functions.

\subsection{Lattice-constrained effective gluon mass}
\label{subsec:mg}

For the thermodynamic input, we employ the lattice equation-of-state
results for pure $SU(3)$ Yang--Mills theory from Ref.~\cite{Borsanyi:2012ve}. The present analysis uses the 29 temperature points in the interval 
$0.70\leq x\leq5$, where $x=T/T_c$. Since the thermodynamic quantities 
considered below are expressed in units of $T_c$, no particular choice of 
the absolute temperature scale is required. This extended range is adopted 
to constrain the high-temperature behavior of the effective mass 
parametrization. Although the parametrization is determined over the full 
interval, the thermodynamic results presented below are mainly shown for 
$0.70\leq x\leq3$, where the nonperturbative effects associated with the 
deconfinement transition are most prominent. 

It is important to clarify the interpretation of the quasiparticle description
in the subcritical region. In pure $SU(3)$ Yang--Mills theory, the physical
excitations below $T_c$ are color-singlet states, with glueballs providing the
dominant low-temperature degrees of freedom
\cite{Borsanyi:2012ve,Buisseret:2010mop}. Therefore, the quasiparticle
representation adopted here is not intended to provide a microscopic
description of the confined-phase spectrum. Instead, it serves as an
effective thermodynamic mapping between the lattice equation of state and a
medium-modified gluonic excitation picture. 
Similar quasigluon-based descriptions have been extended into the confined
region, where confinement effects are incorporated through a strong
suppression or modification of colored thermal excitations
\cite{Heffner:2012sx,Filinov:2014dda,Islam:2021qwh,Shen:2026wvb,Kang:2022jbg}. Within the
present framework, the effective mass $m_g(T)$ should therefore be understood
as a thermodynamic parameter rather than a pole mass of a propagating gluon.
In particular, for $T<T_c$, the large value of $m_g/T$ does not indicate the
existence of heavy physical gluon states; instead, it represents the strong
suppression of thermally excited gluonic modes required to reproduce the small
entropy density observed in lattice calculations.

As shown in Sec.~\ref{sec:II}, the background contribution cancels from
the enthalpy density. At vanishing chemical potential, the entropy density
therefore depends only on the quasiparticle sector. We consequently define
the effective mass at each lattice temperature $T_i$ by matching the
quasiparticle entropy to the lattice result,
\begin{equation}
\left.
\frac{S_{\rm id}(T,m_g)}{T^3}
\right|_{T=T_i}
=
\left(
\frac{S}{T^3}
\right)^{\rm lat}_{i}.
\label{eq:entropy_matching}
\end{equation}
For the lattice data of Ref.~\cite{Borsanyi:2012ve}, the entropy density
may equivalently be reconstructed from the tabulated pressure and
interaction measure through
\begin{equation}
\frac{S(T)}{T^3}
=
\frac{I(T)+4P(T)}{T^4},
\end{equation}
where $I(T)=E(T)-3P(T)$.
Equation~\eqref{eq:entropy_matching} then yields the discrete set
\begin{equation}
q_i
=
\left.
\frac{m_g(T)}{T}
\right|_{T=T_i}.
\label{eq:qi}
\end{equation}
These values characterize the effective thermal suppression scale required
by the quasiparticle ansatz to reproduce the lattice entropy density. Their
interpretation as a physical gluon mass is appropriate only within the
limitations of the effective description discussed above.

A continuous representation of these discrete values is required for the
subsequent thermodynamic reconstruction because the consistency relation for
$B(T)$ involves the temperature derivative of the effective mass. In
particular, a smooth parametrization is necessary to evaluate the background
contribution reliably while retaining the rapid variation of the lattice-
constrained effective mass near $T_c$. The parametrization must capture
the rapid variation around the deconfinement temperature together with the
much smoother behavior away from the transition, while remaining suitable
for numerical differentiation and integration.

For this purpose, we introduce
\begin{equation}
u=\frac{x-0.840}{0.140},
\qquad
z=700.000(x-1),
\label{eq:uz}
\end{equation}
and the switching function
\begin{equation}
{\cal W}(x)
=
\frac{1}
{1+e^{\left(z+1.054\right)}}.
\label{eq:switch}
\end{equation}
The dimensionless effective mass is parametrized as
\begin{equation}
\frac{m_g(T)}{T}
=
{\cal W}(x)L(u)
+
\left[1-{\cal W}(x)\right]H(x),
\label{eq:mg_fit}
\end{equation}
with
\begin{align}
L(u)
={}&
8.740
-1.452u
-0.002u^2
+0.112u^3
\nonumber\\
&
+0.025u^4
-0.619u^5
-0.058u^6
+0.290u^7 ,
\label{eq:L_branch}
\end{align}
and
\begin{align}
H(x)
={}&
0.019
+1.008\,e^{-0.054(x-1)}
+0.805\,e^{-0.976(x-1)}
\nonumber\\
&
+0.877\,e^{-3.771(x-1)}
+0.633\,e^{-14.005(x-1)}
\nonumber\\
&
+0.517\,e^{-48.973(x-1)} .
\label{eq:H_branch}
\end{align}
The first branch describes the subcritical and near-transition region, whereas
the second accounts for the smoother evolution at higher temperatures.
The smooth switching function introduced above is employed to provide
a numerically stable interpolation of the discrete entropy-derived effective mass values in the narrow temperature region around $T_c$. This interpolation is particularly useful because the thermodynamic consistency relation involves the
temperature derivative $d m_g(T)/dT$ when reconstructing the background
contribution. The resulting smooth parametrization should thus be interpreted as an effective representation of the lattice-constrained quasiparticle mass, capturing its rapid temperature dependence near $T_c$ while avoiding numerical instabilities in subsequent thermodynamic calculations.

The parameters in Eqs.~\eqref{eq:switch}--\eqref{eq:H_branch} are obtained
from an unweighted least-squares fit to the values extracted from
Eq.~\eqref{eq:entropy_matching},
\begin{equation}
{\cal L}
=
\sum_i
\left[
q_{\rm fit}(x_i)-q_i
\right]^2,
\qquad
q_{\rm fit}(x)
\equiv
\frac{m_g(T)}{T}.
\label{eq:mass_fit_objective}
\end{equation}
Over the full fitting interval $0.70\leq T/T_c\leq5$, the parametrization 
yields a root-mean-square (RMS) deviation of approximately
\begin{equation}
\Delta_{\rm RMS}\left(\frac{m_g}{T}\right)
\simeq
4.96\times10^{-4},
\end{equation}
while the maximum relative deviation remains below approximately
$0.07\%$. Substituting the fitted function back into the quasiparticle
entropy gives an RMS deviation of approximately
$8.2\times10^{-4}$ in $S/T^3$, demonstrating that replacing the discrete
values by the analytic parametrization introduces only a very small
numerical error.

\begin{figure}
\centering
\includegraphics[width=1\linewidth]{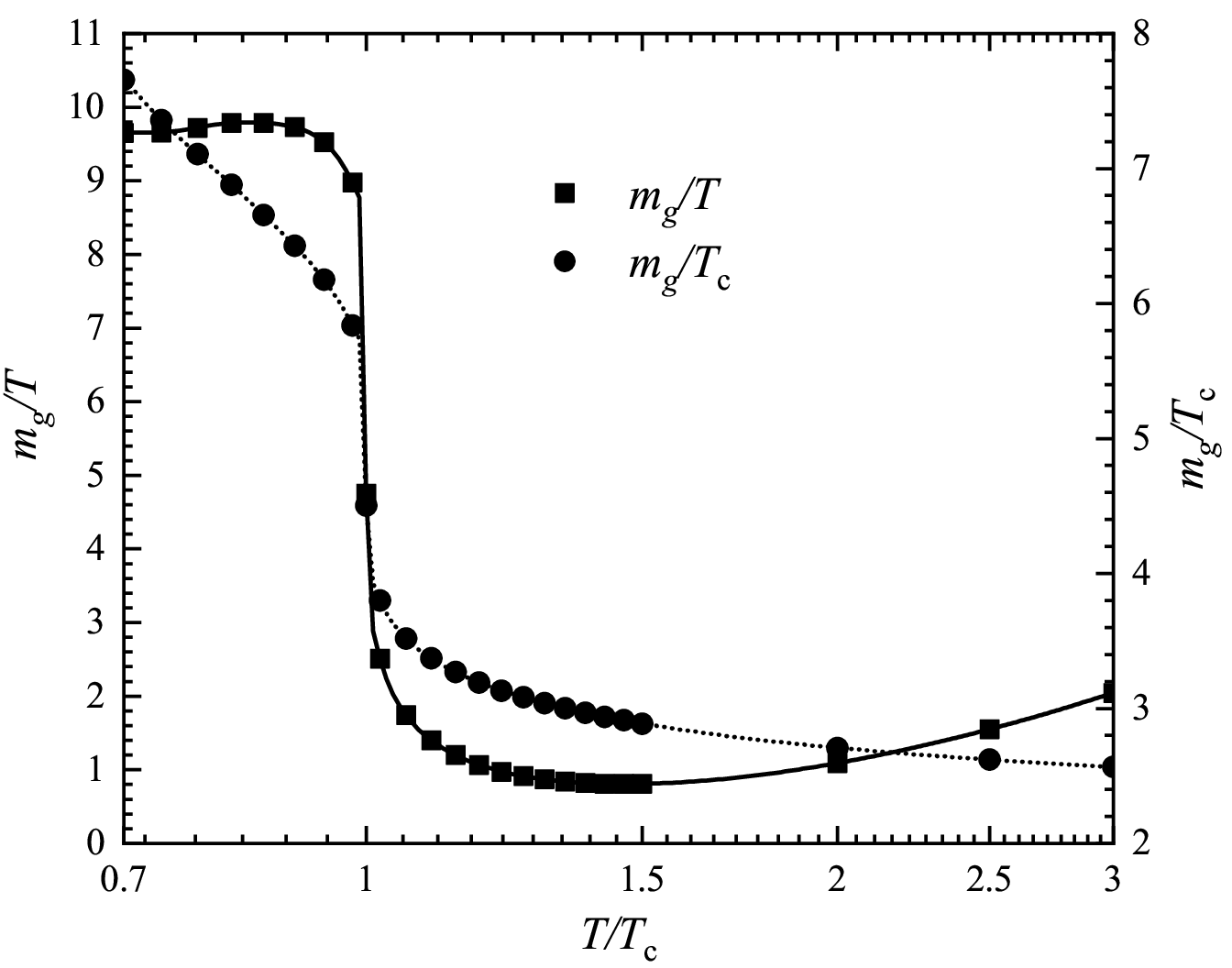}
\caption{
Temperature dependence of the effective gluon quasiparticle mass 
in units of $T$ and $T_c$. The square and circular symbols denote the values 
of $m_g/T$ and $m_g/T_c$, respectively, obtained from entropy matching to the 
lattice equation of state. The solid curves correspond to the parametrization 
given in Eq.~\eqref{eq:mg_fit}, fitted over $0.70\leq T/T_c\leq5$. The shown 
range is restricted to $T/T_c\leq3$ for consistency with the thermodynamic 
analysis.
}
\label{fig:MTTc}
\end{figure}

Figure~\ref{fig:MTTc} shows the effective quasiparticle mass extracted by matching the quasiparticle entropy density
to the corresponding lattice results. Below $T_c$, the relatively
large values of $m_g/T$ indicate a strong suppression of thermally
active gluonic modes. 
This behavior should not be interpreted
as evidence for freely propagating massive gluons in the confined phase.
Instead, the small lattice entropy of the glueball-dominated region is
mapped by Eq.~\eqref{eq:entropy_matching} onto a large effective
quasiparticle mass. In this sense, $m_g/T$ below $T_c$ acts primarily as
an effective thermal suppression scale. Similar quasigluon descriptions
of the confined region have been considered in approaches where
confinement strongly modifies the colored excitation spectrum
\cite{Heffner:2012sx,Islam:2021qwh,Lo:2013hla,Filinov:2014dda,Kang:2022jbg}, although the microscopic implementations of confinement in those models
differ from the thermodynamic parametrization employed here.

In the vicinity of $T_c$, $m_g/T$ changes rapidly over a narrow
temperature interval, reflecting the pronounced
temperature dependence of the lattice entropy near the deconfinement
transition. Above $T_c$, its evolution becomes
considerably smoother: $m_g/T$ decreases with increasing
temperature, with the rate of decrease becoming progressively weaker
at higher temperatures.  
The behavior of $m_g/T_c$ differs from that of $m_g/T$ because
$m_g/T_c=xm_g/T$.
Consequently, a decrease of $m_g/T$ does not necessarily imply a decrease
of the mass measured relative to the fixed scale $T_c$. The quantity
$m_g/T$ directly controls the Bose--Einstein occupation factors and is
therefore the more relevant measure of thermal suppression, whereas
$m_g/T_c$ characterizes the corresponding effective dimensional scale. 
Over the full interval $0.70\leq T/T_c\leq3$, the strongest variation of
$m_g/T$ is concentrated around the deconfinement region. At several times
$T_c$ the temperature dependence is much weaker, but this regime should
not be identified with the strict asymptotic limit of Yang--Mills theory.
Indeed, lattice calculations show that significant deviations from the
ideal-gas limit persist even at several times $T_c$
\cite{Boyd:1996bx,Borsanyi:2012ve}.

Once $m_g(T)$ has been determined from the lattice entropy density, the temperature-dependent part of the quasiparticle model is fixed. The solution of the thermodynamic-consistency condition for the background contribution contains one integration constant, which may be specified by fixing the pressure at a reference temperature. We choose $T_0=T_c$ and require the quasiparticle-model pressure to reproduce the corresponding lattice value, $P(T_c)=P_{\mathrm{lat}}(T_c)$. According to Eq.~\eqref{eq:P_qpm}, this condition yields $B(T_c)=P_{\rm id}(T_c,m_g(T_c))
-P_{\rm lat}(T_c)$. For the lattice input adopted in this work, $P_{\rm lat}(T_c)/T_c^4=0.0222$, $m_g(T_c)/T_c\simeq4.591$, 
while the corresponding quasiparticle contribution is $P_{\rm id}(T_c,m_g(T_c))/T_c^4 \simeq0.1467$. 
It therefore follows that 
\begin{align}
B(T_0) = B(T_c)\simeq0.1245 T_c^4.
\end{align}
This quantity does not constitute an additional parameter governing the temperature dependence of the model; rather, it is the single thermodynamic normalization constant required to specify the solution of Eq.~\eqref{eq:Bstar}. Once $m_g(T)$ and $B(T_c)$ are fixed, the function $B(T)$, and consequently the full equation of state, are uniquely determined.

\subsection{Equation of state and interaction measure}
\label{subsec:eos}

We next use the lattice-constrained effective mass
$m_g(T)$ in Eq.~\eqref{eq:mg_fit} to reconstruct the thermodynamic
equation of state. The pressure and energy density are evaluated from
Eqs.~\eqref{eq:P_qpm} and \eqref{eq:E_qpm}, respectively, with the
background contribution determined from Eq.~\eqref{eq:Bstar}. Since
the entropy density is used as the primary input to constrain $m_g(T)$,
its reproduction provides a consistency check on the parametrized
effective mass, whereas the pressure and energy density are subsequently reconstructed from the thermodynamic consistency relations after fixing the single pressure normalization condition at $T_c$.

\begin{figure}
\centering
\includegraphics[width=1\linewidth]{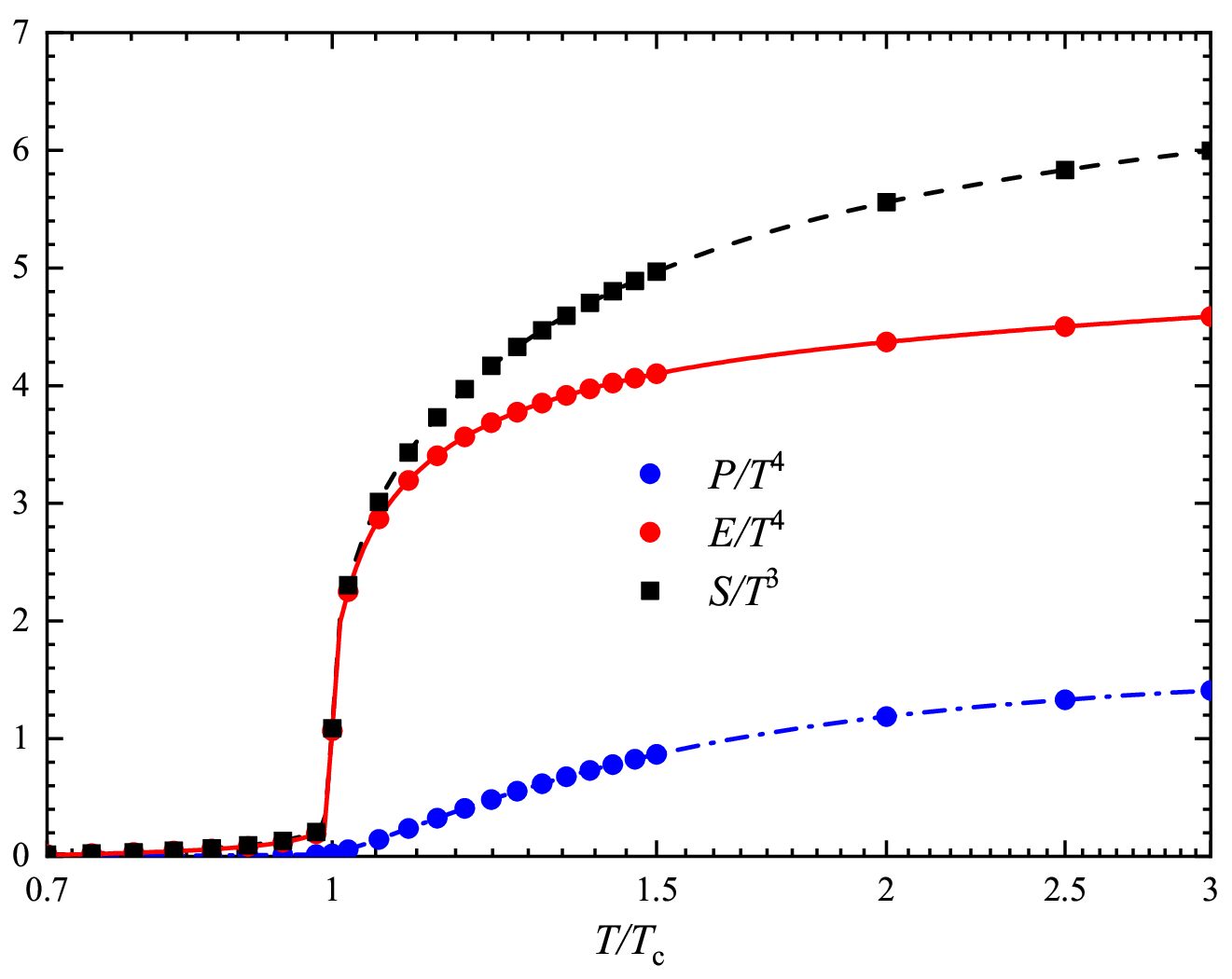}
\caption{
Temperature dependence of the normalized pressure $P/T^4$, energy
density $E/T^4$, and entropy density $S/T^3$. The curves denote the
thermodynamically consistent quasiparticle results, while the symbols
represent the corresponding lattice data~\cite{Borsanyi:2012ve}.
}
\label{fig:PEST}
\end{figure}

Figure~\ref{fig:PEST} compares the reconstructed equation of state with
the corresponding lattice results. All three thermodynamic quantities
remain strongly suppressed below $T_c$ and change sharply across the
deconfinement region. In particular, $E/T^4$ and $S/T^3$ exhibit a much
more pronounced variation near $T_c$, whereas $P/T^4$ evolves more
smoothly. This behavior reflects the pronounced temperature
dependence of the thermodynamic properties of pure $SU(3)$ Yang--Mills
matter near the deconfinement transition. 
Above $T_c$, the temperature dependence becomes
progressively smoother as the system moves away from the transition
region.

The quasiparticle results reproduce the lattice data well throughout
the temperature interval considered. This agreement reflects the internal consistency of the thermodynamic reconstruction, in which the pressure and energy density are obtained from the lattice-constrained effective spectrum and the thermodynamic consistency condition without additional temperature-dependent fitting functions. 
The temperature dependence of $m_g(T)$ is determined from the lattice
entropy density, and only the pressure at $T_c$ is used to fix the
integration constant associated with the background contribution.
Once these quantities are specified, $P(T)$ and $E(T)$ follow from
thermodynamic consistency. Their agreement with lattice data confirms that the lattice-constrained effective quasiparticle framework consistently reproduces the equation of state once the entropy constraint and thermodynamic consistency condition are imposed.

A particularly sensitive measure of the deviation from conformal
thermodynamics is the interaction measure, $I(T)=E(T)-3P(T)$. 
Within the present quasiparticle formulation, it can be expressed as
\begin{equation}
I(T)
=
E_{\rm id}(T,m_g)
-
3P_{\rm id}(T,m_g)
+
4B(T).
\label{eq:I_decomposition}
\end{equation}
The first term in Eq.~\eqref{eq:I_decomposition} arises from the
nonzero effective quasiparticle mass, while the second originates from
the background contribution required by thermodynamic consistency.
These terms should not be regarded as independent sources of
nonconformality, since $B(T)$ is itself determined by the temperature
dependence of $m_g(T)$ through Eq.~\eqref{eq:Bstar}. Together, they
encode the deviation of the thermodynamic behavior of the system from the conformal limit.

\begin{figure}
\centering
\includegraphics[width=1\linewidth]{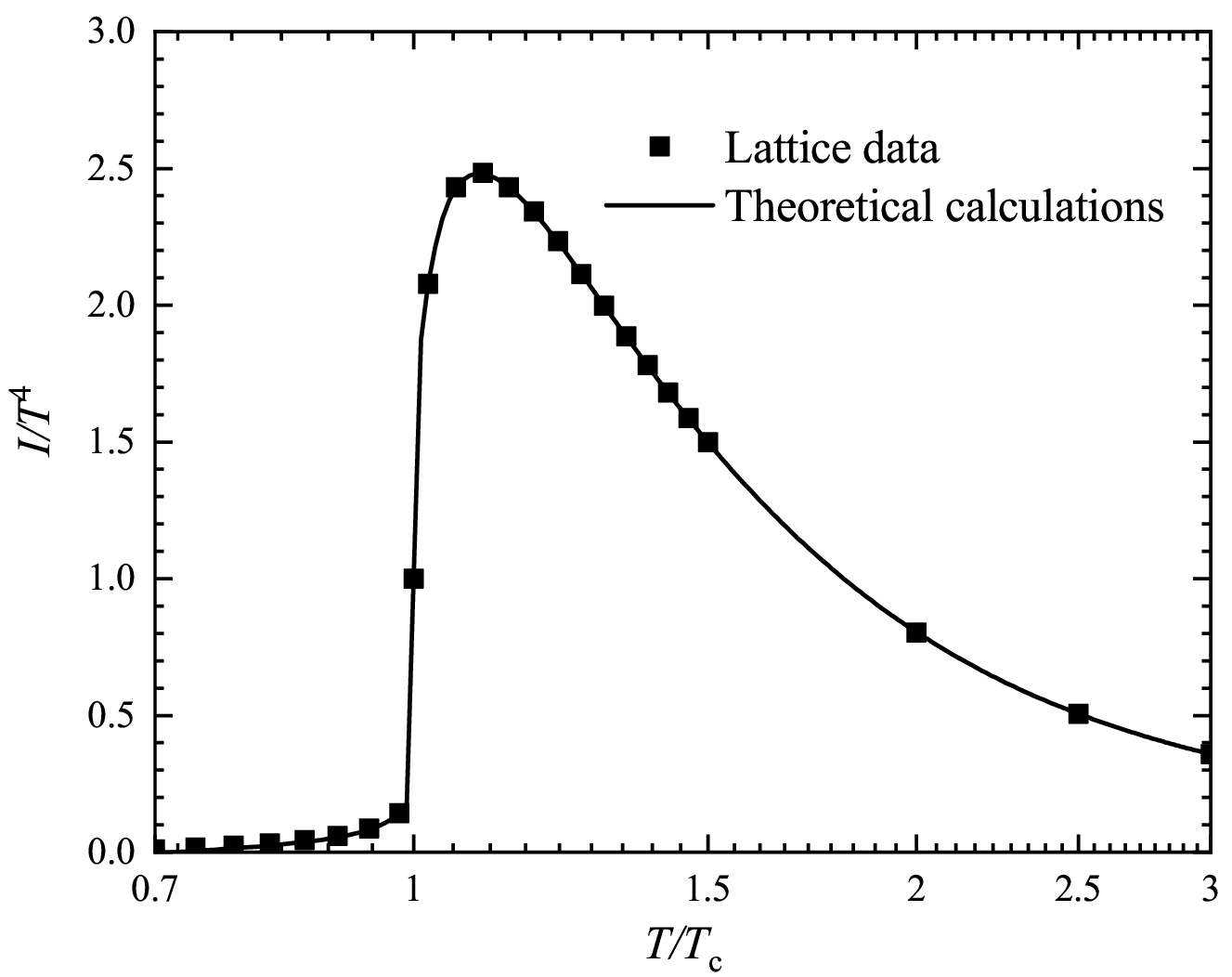}
\caption{
Normalized interaction measure $I/T^4$ as a function of $T/T_c$.
The curve denotes the thermodynamically consistent quasiparticle
result, while the symbols represent the corresponding lattice data~\cite{Borsanyi:2012ve}.
}
\label{fig:IT4}
\end{figure}

As shown in Fig.~\ref{fig:IT4}, the interaction measure exhibits a rapid
increase across the deconfinement region and reaches a pronounced maximum
slightly above the critical temperature. In the present calculation, the
peak is located at
\begin{equation}
\frac{T_{\rm peak}}{T_c}\simeq 1.09,
\qquad
\left(\frac{I}{T^4}\right)_{\rm peak}\simeq 2.48 .
\end{equation}
Although the deconfinement transition takes place at $T_c$, the maximum of the interaction measure does not necessarily coincide with the transition temperature. This is because $I/T^4$ characterizes the deviation from
conformal behavior rather than the phase transition point itself. In the
quasiparticle description, the enhancement of the interaction measure is
mainly associated with the rapid temperature variation of the effective
gluon mass $m_g(T)$ and the corresponding background contribution $B(T)$
required by thermodynamic consistency. Around
$T\simeq1.09T_c$, the effective gluonic excitation spectrum undergoes the most rapid change, resulting in the strongest breaking of conformal
symmetry and hence the maximum of $I/T^4$.

At higher temperatures, $I/T^4$ decreases steadily as the temperature
increases, signaling a gradual evolution toward approximately conformal
thermodynamics. Nevertheless, it remains finite throughout the range
$T/T_c\leq3$, indicating that appreciable interaction effects persist even
at several times the critical temperature. The system therefore approaches
the conformal and Stefan--Boltzmann limits only gradually rather than
immediately after deconfinement.

\subsection{Thermodynamic quantities per thermal quasigluonic excitation and background contribution}
\label{subsec:pergluon}

The quasiparticle framework also allows the bulk thermodynamic quantities
to be normalized by the number density of thermally populated gluonic
excitations. For the dispersion relation introduced above, the thermal
quasiparticle number density is defined as
\begin{equation}
n_g(T)
=
\frac{d_g}{2\pi^2}
\int_0^\infty
\frac{k^2 dk}
{e^{\omega(k,T)/T}-1}.
\label{eq:ng}
\end{equation}
This definition is analogous to the thermal occupation number used in
massive quasiparticle descriptions of gluonic matter
\cite{Peshier:1995ty,Gorenstein:1995vm}. Since gluon quasiparticle
number is not associated with a conserved charge, however, $n_g(T)$
should not be interpreted as a conserved particle-number density.
Instead, it characterizes the abundance of thermal gluonic excitations within the effective quasiparticle description. 

\begin{figure}
\centering
\includegraphics[width=1\linewidth]{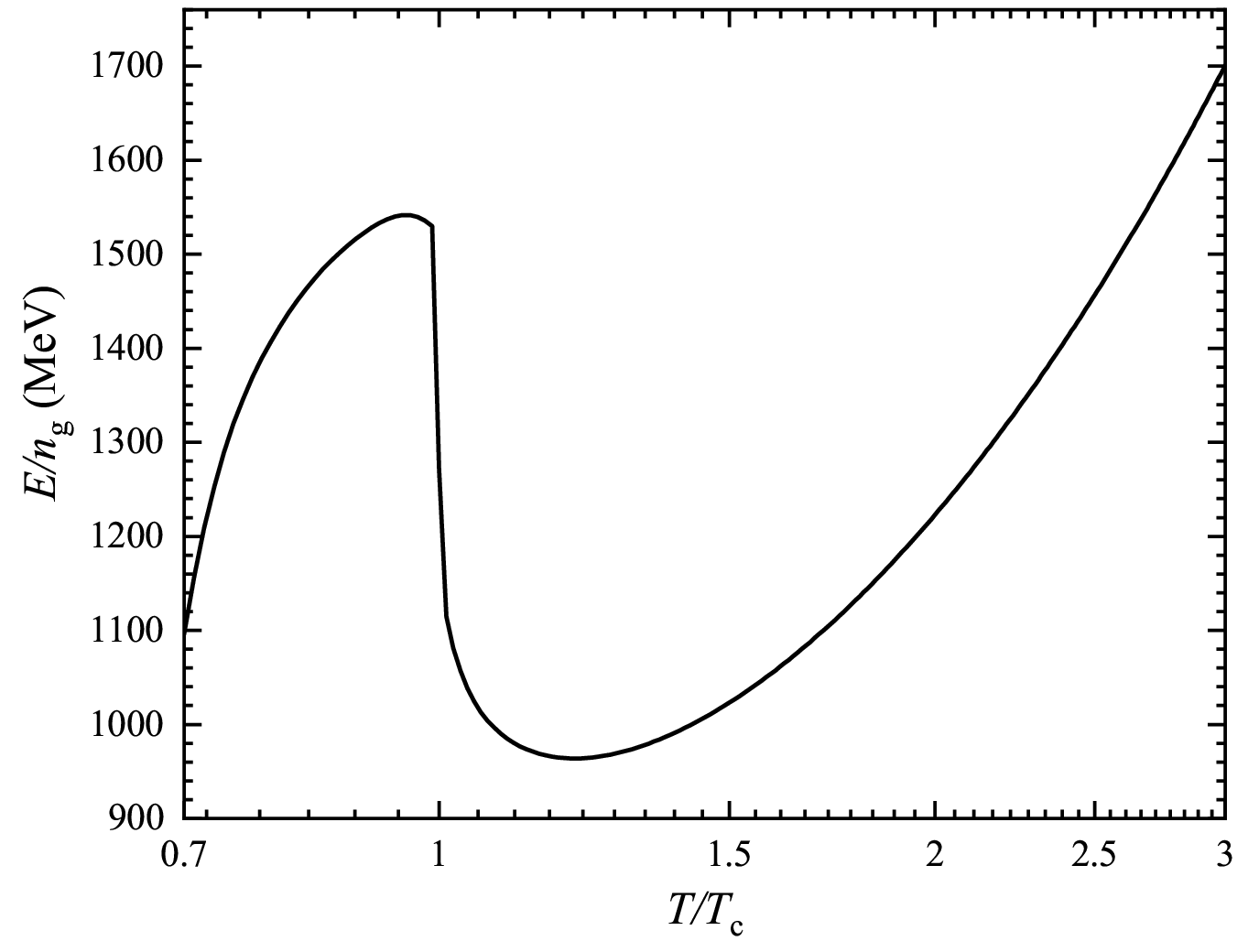}
\caption{
Average energy per thermal quasigluonic excitation, $E/n_g$, as a function of
$T/T_c$. The quantity is evaluated within the thermodynamically consistent
quasiparticle framework and reflects the evolution of the effective
thermal gluonic excitations across the deconfinement transition. 
}
\label{fig:Eng}
\end{figure}

As shown in Fig.~\ref{fig:Eng}, $E/n_g$ exhibits a pronounced
nonmonotonic temperature dependence. On the low-temperature side, it
initially increases as $T$ approaches $T_c$. This behavior should be
interpreted with caution because, below $T_c$, the quasiparticle
description represents an effective continuation of the lattice
thermodynamics rather than the physical glueball spectrum. In this
region, $n_g$ is strongly suppressed by the large value of $m_g/T$, and
therefore the ratio normalized by the quasiparticle density becomes
particularly sensitive to the effective parametrization. 
Across the deconfinement region, $E/n_g$ decreases sharply and reaches a
minimum slightly above the critical temperature. In the present
calculation, the minimum is located at
\begin{equation}
\frac{T_{\rm min}}{T_c}\simeq1.21,
\qquad
\left(\frac{E}{n_g}\right)_{\rm min}
\simeq 963.83~{\rm MeV}.
\end{equation}
This minimum appears after the peak of the interaction measure and the
background contribution, which occur around $T\simeq1.09T_c$. The small
separation between these characteristic temperatures indicates that the
maximum violation of conformal symmetry and the minimum average energy
carried by each thermal excitation are related but not identical
phenomena.

The nonmonotonic behavior of $E/n_g$ originates from the different
temperature dependences of the total energy density and the thermal
quasiparticle population. In particular, near the deconfinement region,
the rapid decrease of the effective gluon mass leads to a rapid increase
of the available gluonic quasiparticle states. Consequently, the
increase of $n_g$ becomes faster than that of the total energy density,
resulting in a reduction of the average energy per excitation. The
minimum therefore reflects the rapid reorganization of the effective
gluonic degrees of freedom in the same temperature interval where
$m_g(T)$ and the bulk thermodynamic quantities exhibit their strongest
temperature dependence.

At higher temperatures, $E/n_g$ increases again. This increase should
not be interpreted directly as a signal of stronger interactions, since
$E/n_g$ is a dimensionful quantity whose characteristic scale naturally
grows with temperature. For comparison, in the ideal gas limit of
massless gluons,
\begin{equation}
n_g^{\rm SB}
=
\frac{d_g\zeta(3)}{\pi^2}T^3,
\qquad
E^{\rm SB}
=
\frac{d_g\pi^2}{30}T^4 ,
\end{equation}
which gives
\begin{equation}
\frac{E^{\rm SB}}{n_g^{\rm SB}}
=
\frac{\pi^4}{30\zeta(3)}T
\simeq2.701T .
\label{eq:Eperg_SB}
\end{equation}
Therefore, an approximately linear increase of the energy carried by
each excitation is expected even in the noninteracting high-temperature
limit. The deviation of the present result from this Stefan--Boltzmann
scaling contains information about the finite quasiparticle mass and the
residual contribution from the thermodynamic background term.

\begin{figure}
\centering
\includegraphics[width=1\linewidth]{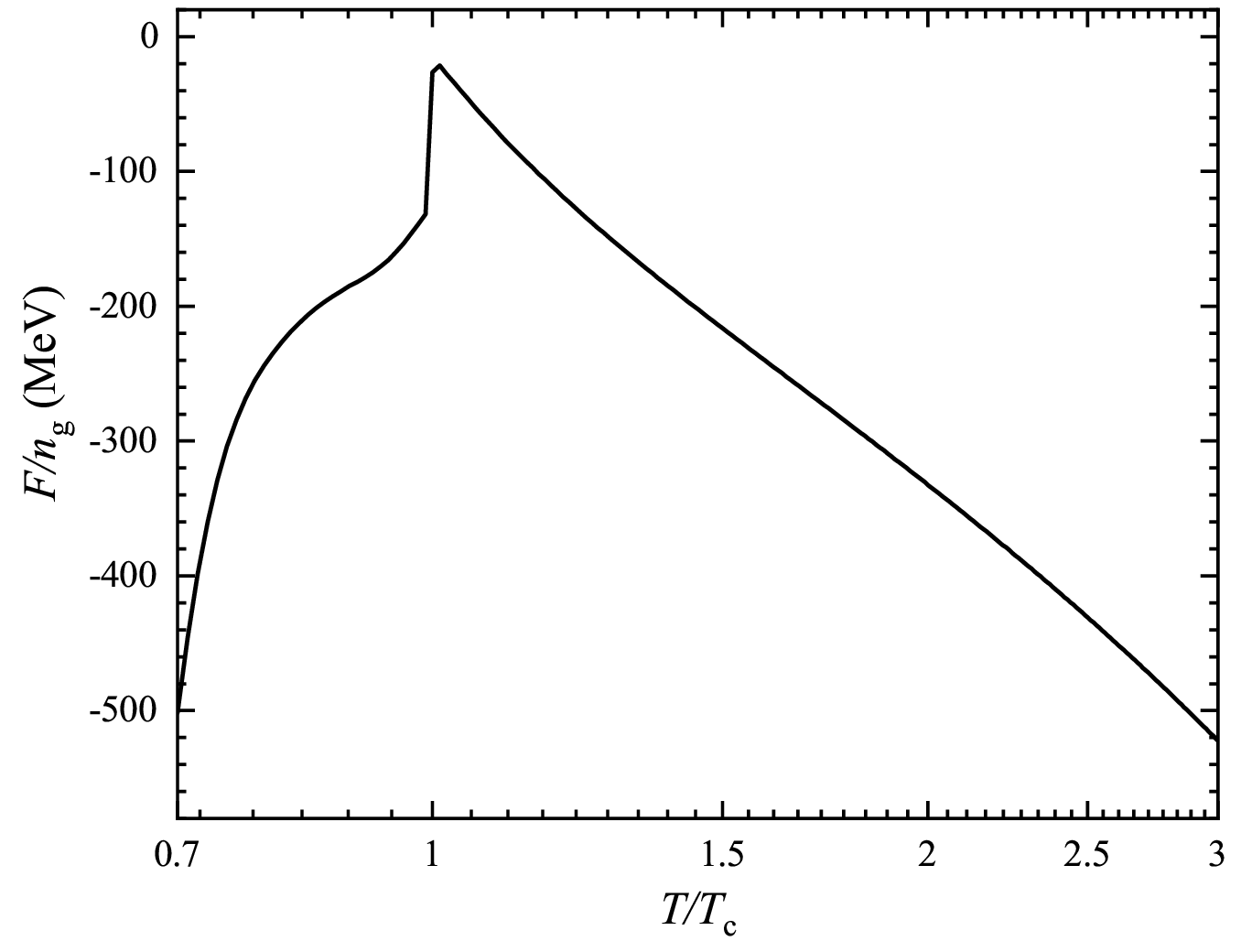}
\caption{
Average free energy per thermal quasigluon, $F/n_g$, as a
function of $T/T_c$. Together with the average energy per gluon, this
quantity provides complementary information on the thermodynamic
properties of the effective gluonic excitations and their evolution across
the deconfinement transition. 
}
\label{fig:Fng}
\end{figure}

At vanishing chemical potential, the thermodynamic relation
$E+P=TS$ implies that the Helmholtz free-energy density satisfies
\begin{equation}
F(T)=E(T)-TS(T)=-P(T).
\end{equation}
Figure~\ref{fig:Fng} shows that $F/n_g$ remains negative throughout
the displayed temperature interval, as expected from the positive
pressure. Below $T_c$, its magnitude decreases as the transition is
approached. A particularly rapid variation occurs around $T_c$, where
the thermal quasiparticle population changes strongly while the
pressure evolves comparatively smoothly. Consequently, the magnitude
of $P/n_g$ is strongly reduced across this region.

Above the transition, $F/n_g$ becomes progressively more negative as
the temperature increases. As in the case of $E/n_g$, part of this
high-temperature behavior follows simply from dimensional scaling. In
the massless Stefan--Boltzmann limit,
\begin{equation}
P^{\rm SB}
=
\frac{d_g\pi^2}{90}T^4,
\end{equation}
and hence
\begin{equation}
\frac{F^{\rm SB}}{n_g^{\rm SB}}
=
-\frac{P^{\rm SB}}{n_g^{\rm SB}}
=-\frac{\pi^4}{90\zeta(3)}T
\simeq
-0.900T.
\label{eq:Fperg_SB}
\end{equation}
The increasingly negative $F/n_g$ at high temperature is therefore
qualitatively consistent with the expected thermal scaling, while the
difference from the massless result reflects the remaining medium
effects encoded in $m_g(T)$ and $B(T)$.

\begin{figure}
\centering
\includegraphics[width=1\linewidth]{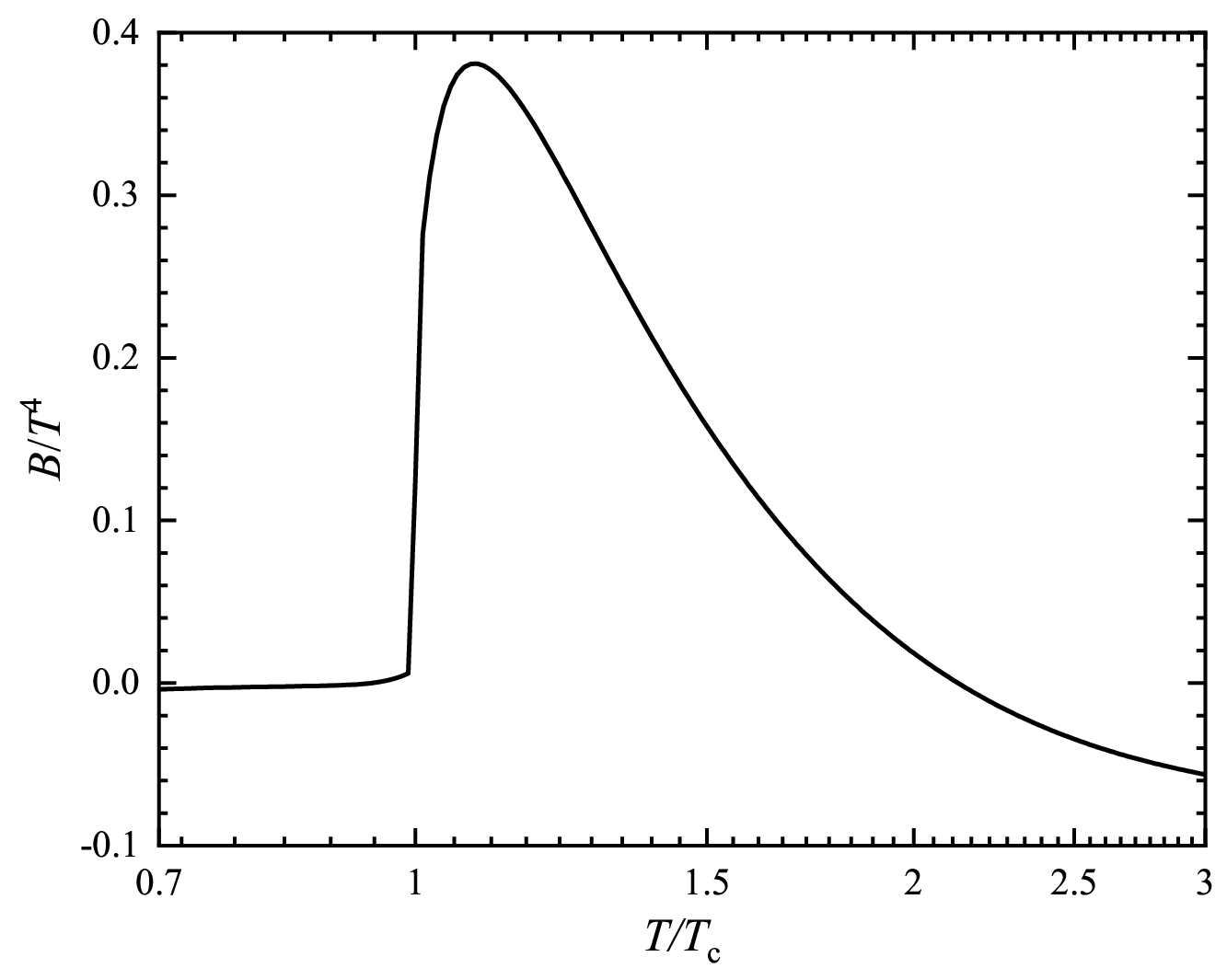}
\caption{
Temperature dependence of the background contribution normalized by
the critical temperature, $B(T)/T^4$.
}
\label{fig:BT4}
\end{figure}

The background contribution required by thermodynamic consistency is
shown separately in Fig.~\ref{fig:BT4}. In quasiparticle models with
a medium-dependent dispersion relation, such a contribution is required
to compensate for the explicit temperature dependence of the effective
mass and to preserve the standard thermodynamic identities
\cite{Gorenstein:1995vm,Peshier:1995ty,Peshier:1999ww,
Schneider:2001nf}. In the present construction, $B(T)$ is not fitted
independently to the lattice equation of state. Once $m_g(T)$ has been
determined and the boundary value $B(T_c)$ has been fixed from the
pressure normalization, its entire temperature dependence follows from
the thermodynamic-consistency condition.

Figure~\ref{fig:BT4} shows that the normalized background contribution
$B(T)/T^4$ exhibits a pronounced structure in the vicinity of the
deconfinement region. It increases rapidly around $T_c$, reaches a broad
maximum slightly above the transition temperature, and gradually
decreases as the temperature is raised further. Interestingly, the
maximum of $B(T)/T^4$ is found at the same temperature as the peak of the
interaction measure, namely
$T_{\rm peak}\simeq1.09T_c$. This behavior is directly related to the temperature dependence of the
effective quasiparticle mass. In particular, the thermodynamic-consistency
condition given in Eq.~\eqref{eq:dB_integral} connects the evolution of
$B(T)$ to the temperature derivative of the quasiparticle mass.
Therefore, the rapid variation of the effective spectrum near $T_c$
naturally generates a strong background contribution in this region.

For the normalization adopted in this work, $B(T)/T^4$ becomes slightly
negative at sufficiently high temperatures within the displayed range.
This sign change does not indicate a thermodynamic instability, nor
should the sign of $B(T)$ alone be interpreted as a direct measure of
the interaction strength. The consistency equation determines the
temperature evolution of the background contribution, whereas its
absolute normalization is fixed by the boundary condition imposed on
the pressure. The physical effects of $B(T)$ enter through its
contributions $-B(T)$ and $+B(T)$ to the pressure and energy density,
respectively.

The normalization $B(T)/T^4$ directly measures the relative importance
of the background contribution compared with the dominant thermal scale.
The decrease of this quantity at high temperatures therefore indicates
that the background term becomes progressively less significant relative
to the quasiparticle thermal contribution, which is consistent with the
expected approach toward the asymptotic high-temperature regime.

\subsection{Thermodynamic response functions}
\label{subsec:response}

Thermodynamic response functions provide a sensitive probe of the rapid
variation of the equation of state near the deconfinement transition.
At vanishing chemical potential, the squared speed of sound is given by
\begin{equation}
c_s^2
=\frac{dP}{dE}
=\frac{dP/dT}{dE/dT}.
\label{eq:cs2}
\end{equation}
A small value of $c_s^2$ corresponds to a soft equation of state,
whereas the conformal limit in three spatial dimensions is
$c_s^2=1/3$.

\begin{figure}
\centering
\includegraphics[width=1\linewidth]{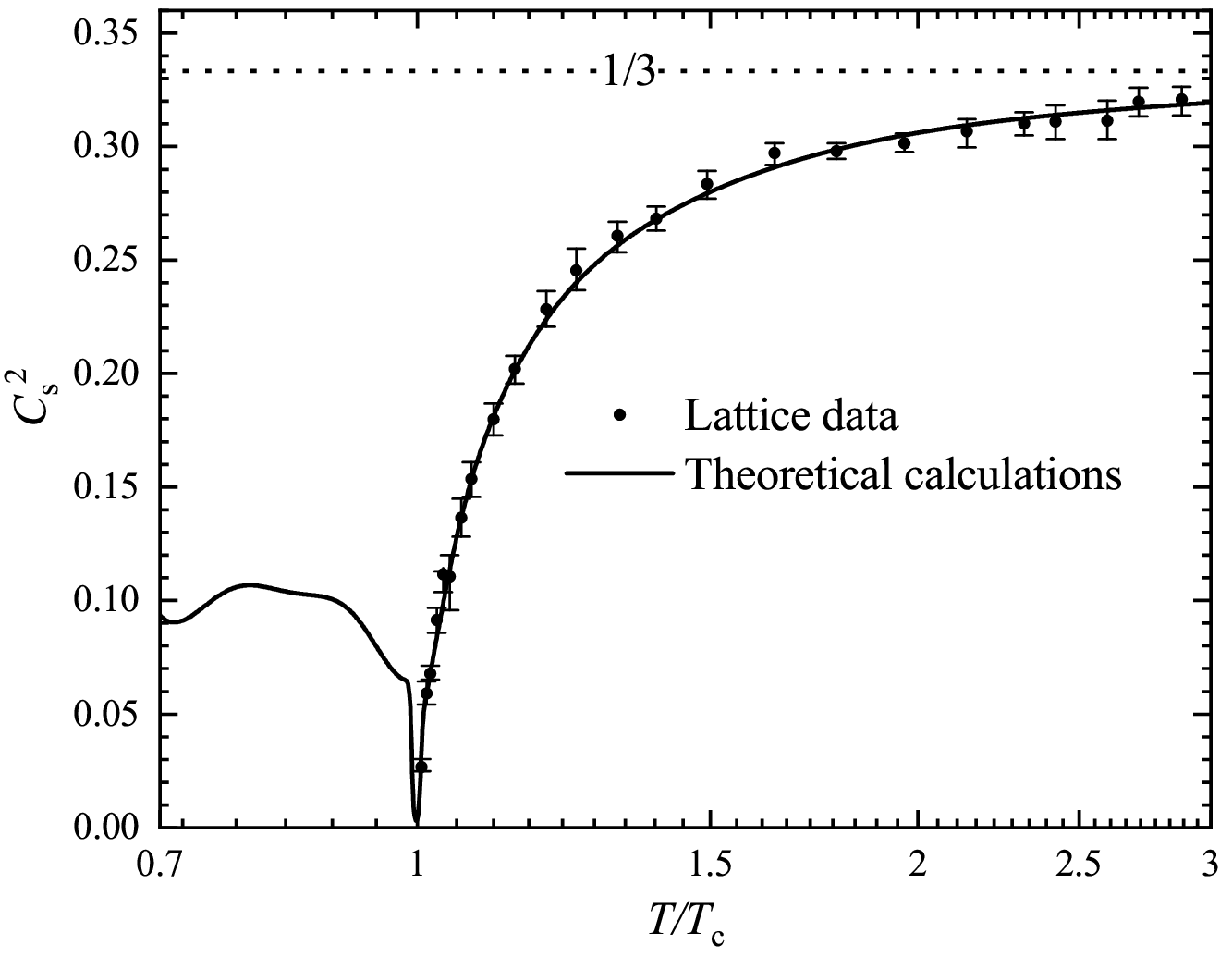}
\caption{
Squared speed of sound $c_s^2$ as a function of $T/T_c$.
The solid curve denotes the thermodynamically consistent quasiparticle
result, while the symbols represent the corresponding lattice data~\cite{Giusti:2025fxu}.
The horizontal dotted line indicates the conformal value $c_s^2=1/3$.
}
\label{fig:Cs2}
\end{figure}

As shown in Fig.~\ref{fig:Cs2}, the quasiparticle calculation
shows good agreement with the temperature dependence
of the squared speed of sound obtained from recent lattice simulations
of pure $SU(3)$ Yang--Mills theory~\cite{Giusti:2025fxu}.
In particular, the calculated curve captures the rapid suppression
of $c_s^2$ in the vicinity of $T_c$,
followed by a smooth increase toward the high-temperature regime.
The good agreement with the lattice data indicates that the
temperature-dependent effective gluon mass provides an effective description of the pronounced softening
of the equation of state near the deconfinement transition.

The pronounced minimum of $c_s^2$ around $T_c$ reflects the strong
softening of the gluonic equation of state. This behavior is closely associated with the rapid variation of the
effective gluon mass and is correlated with the enhancement of the
interaction measure in the same temperature region.
Such a correlation highlights that the nonconformal thermodynamic
behavior near the transition is closely related to the rapid change
of the effective gluon mass and the associated
thermodynamic background contribution~\cite{Gavai:2005dy,Karsch:2008fe,Datta:2009jn}.
For comparison, in QCD matter with dynamical quarks at physical quark masses and vanishing baryon chemical
potential, the transition becomes a crossover and the suppression
of $c_s^2$ is generally less pronounced. The strong suppression of $c_s^2$ near $T_c$ in pure
Yang--Mills matter reflects the pronounced softening of its
equation of state in the deconfinement region. Nevertheless, both pure gluon matter and
quark--gluon plasma approach the conformal limit
$c_s^2=1/3$ at sufficiently high temperatures, while interaction effects can produce deviations from this limit
at finite temperatures~\cite{HotQCD:2014kol,Borsanyi:2013bia,CMS:2024sgx}.

Above the transition region, $c_s^2$ increases rapidly and gradually
approaches the conformal limit~\cite{Itou:2025vcy}.
At $T/T_c\simeq3$, the calculated value
remains slightly below $1/3$, consistent with the nonvanishing
interaction measure in the same temperature range.
This indicates that although the system becomes approximately
conformal at high temperature, interaction effects remain relevant even at temperatures several times $T_c$. The smooth minimum shown in Fig.~\ref{fig:Cs2}
should therefore be interpreted as the thermodynamic softening
captured by the lattice-constrained quasiparticle equation of state,
rather than as an explicit description of the thermodynamic
discontinuities at $T_c$.


A complementary measure of the mechanical response of the gluon plasma is provided by the isentropic bulk modulus,
\begin{equation}
K_S
=
(E+P)c_s^2
 =TSc_s^2.
\label{eq:Ks}
\end{equation}
At vanishing chemical potential, $E+P=TS$, so that $K_S$ combines the
enthalpy scale of the medium with the stiffness encoded in $c_s^2$.
A smaller value of $K_S$ therefore corresponds to a softer and more
compressible system.

\begin{figure}
\centering
\includegraphics[width=1\linewidth]{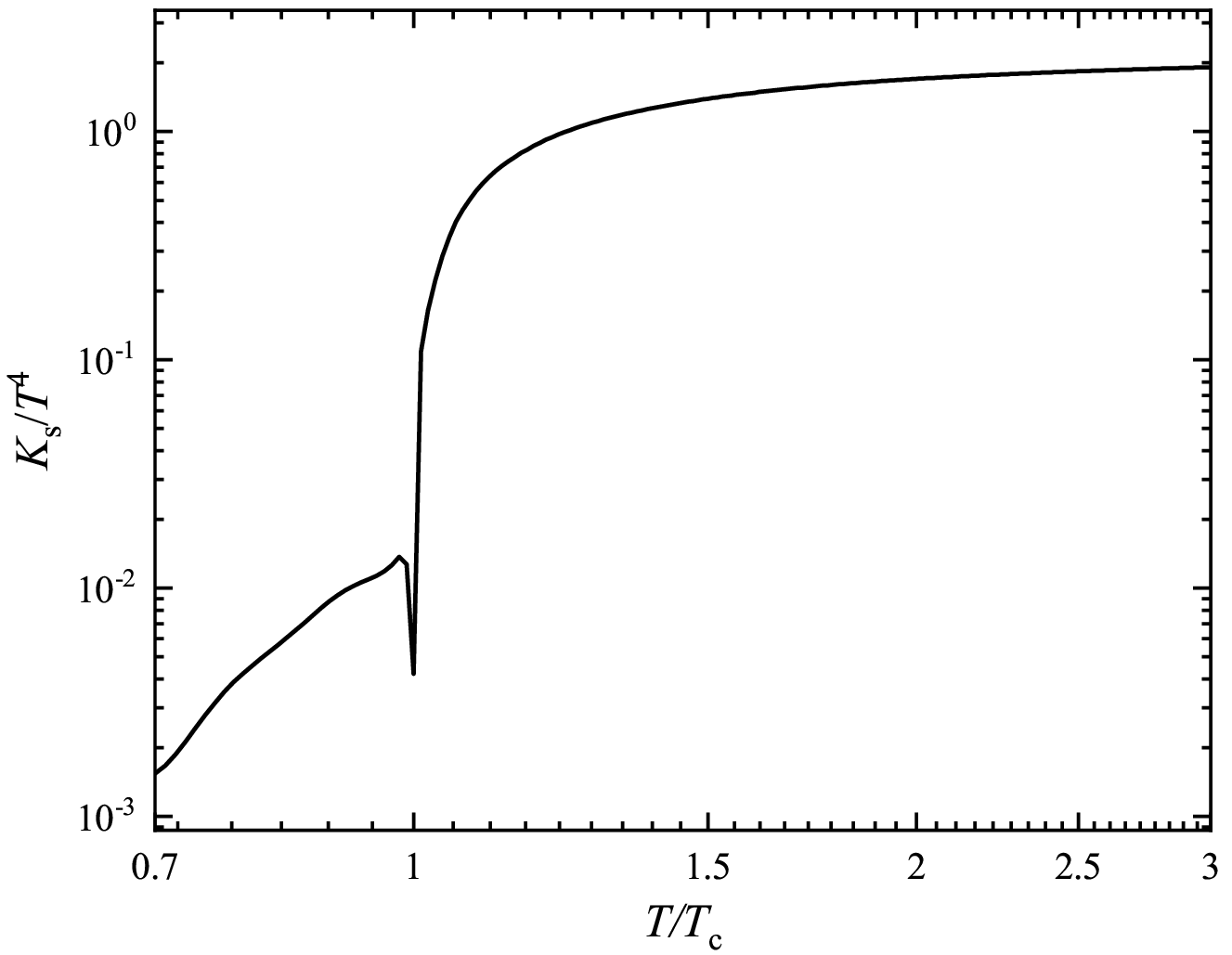}
\caption{
Normalized isentropic bulk modulus $K_S/T^4$ as a function of
$T/T_c$. The vertical axis is shown on a logarithmic scale.
}
\label{fig:Ks}
\end{figure}

Figure~\ref{fig:Ks} shows that $K_S/T^4$ is strongly suppressed near
the deconfinement transition, consistently identifying this region as
the softest part of the equation of state. Immediately above $T_c$,
$K_S/T^4$ increases rapidly as both the enthalpy density and the speed
of sound recover, while its temperature dependence becomes much smoother
at higher temperatures.

In the Stefan--Boltzmann limit of a massless pure-gluon gas,
\begin{equation}
\frac{K_S^{\rm SB}}{T^4}
=
\frac{2d_g\pi^2}{135}
\simeq2.34 .
\label{eq:Ks_SB}
\end{equation}
The high-temperature behavior in Fig.~\ref{fig:Ks} is consistent with
a gradual approach toward this limit, although noticeable deviations
remain at $T/T_c=3$.

Taken together, the interaction measure, squared speed of sound, and
isentropic bulk modulus provide complementary descriptions of the same
thermodynamic evolution. The enhancement of $I/T^4$ quantifies the
strong departure from conformal behavior, while the suppression of
$c_s^2$ and $K_S$ characterizes the associated softening of the
equation of state near the deconfinement transition.


\section{Conclusions}
\label{sec:conclusion}

In this work, we have constructed a thermodynamically consistent
quasiparticle description of finite-temperature pure $SU(3)$ Yang--Mills
matter constrained by lattice thermodynamics. The temperature-dependent
effective gluon mass $m_g(T)$ is determined directly from the lattice
entropy density, while the corresponding background contribution $B(T)$ is
subsequently reconstructed from the thermodynamic consistency condition.
Only one normalization condition, imposed through the pressure at
$T_c$, is required to fix the integration constant of $B(T)$. Once these
quantities are specified, the pressure, energy density, entropy density,
interaction measure, and squared speed of sound follow without introducing
additional temperature-dependent fitting functions. This provides a compact
connection between lattice thermodynamic information and an effective
description of the underlying gluonic excitations.

The resulting thermodynamic observables reproduce the characteristic
features of pure $SU(3)$ Yang--Mills lattice results over the temperature
range considered. The strongest medium modifications occur in the vicinity
of the deconfinement transition, where the effective gluon mass exhibits a
rapid variation, the interaction measure develops a pronounced enhancement,
and the squared speed of sound is strongly reduced. These behaviors
indicate a substantial departure from conformal thermodynamics and a
significant softening of the equation of state near $T_c$. At higher
temperatures, the interaction measure gradually decreases and $c_s^2$
approaches the conformal limit, although the system remains noticeably
different from the Stefan--Boltzmann ideal-gas limit at several times
$T_c$. The energy and free energy per thermal quasigluon, together with the
temperature dependence of the background contribution and the isentropic
bulk modulus, further characterize the evolution of the effective gluonic
degrees of freedom and the mechanical response of the medium.

The quasiparticle mass obtained below $T_c$ should be interpreted as an
effective thermodynamic parameter that encodes the suppression of thermal
gluonic excitations, rather than as the mass of a physical propagating
gluon in the confined phase. Within this interpretation, the present
framework provides a unified description of lattice thermodynamics across
the deconfinement region while retaining a simple quasiparticle
representation. The approach can be further extended to investigate
non-equilibrium properties and transport coefficients, such as shear and
bulk viscosities, and may also provide a useful starting point for studies
of QCD matter with dynamical quarks, finite chemical potentials, or external
fields.

\section*{Acknowledgments}

We thank Wei Shen and Guang-Xiong Peng for useful discussions. 
This work is supported by  
the National Natural Science Foundation of China  (Grant No.~12205093), 
the Natural Science Foundation of Hunan Province (Grant No.~2026JJ50352), and the Scientific Research Foundation of Hunan Provincial Education Department (Grants No.~25B0432 and No.~25C0302).


\bibliographystyle{aapmrev4-2}   
\bibliography{Ref}



\end{document}